\documentclass[letterpaper,twocolumn,english,prl]{revtex4}

\usepackage[T1]{fontenc}
\usepackage[utf8x]{inputenc}
\usepackage{color}
\usepackage{amstext}
\usepackage{stackrel}
\usepackage{graphicx}
\usepackage{esint}

\makeatletter

\@ifundefined{textcolor}{}
{%
 \definecolor{BLACK}{gray}{0}
 \definecolor{WHITE}{gray}{1}
 \definecolor{RED}{rgb}{1,0,0}
 \definecolor{GREEN}{rgb}{0,1,0}
 \definecolor{BLUE}{rgb}{0,0,1}
 \definecolor{CYAN}{cmyk}{1,0,0,0}
 \definecolor{MAGENTA}{cmyk}{0,1,0,0}
 \definecolor{YELLOW}{cmyk}{0,0,1,0}
}

\@ifundefined{definecolor}
 {\usepackage{color}}{}

\makeatother

\usepackage{babel}
\begin{document}

\title{Optimal stochastic restart renders fluctuations in first passage
times universal }

\author{{\normalsize{}Shlomi Reuveni$^{\dagger}$}}

\affiliation{\noindent \textit{$^{\dagger}$Department of Systems Biology, Harvard
Medical School, 200 Longwood Avenue, Boston, Massachusetts 02115,
USA.}}
\begin{abstract}
{\normalsize{}Stochastic restart may drastically reduce the expected
run time of a computer algorithm, expedite the completion of a complex
search process, or increase the turnover rate of an enzymatic reaction.
These diverse first-passage-time (FPT) processes seem to have very
little in common but it is actually quite the other way around. Here
we show that the relative standard deviation associated with the FPT
of an optimally restarted process, i.e., one that is restarted at
a (non-zero) rate which brings the mean FPT to a minimum, is always
unity. We interpret, further generalize, and discuss this finding
and the implications arising from it. }{\normalsize \par}
\end{abstract}
\maketitle
Stopping a process in its midst—only to start it all over again—may
prolong, leave unchanged, or even shorten the time taken for its completion.
Among these three possibilities the latter is particularly interesting
as it suggests that restart can be used to expedite the completion
of complex processes involving strong elements of chance. This observation
has long been made in the field of computer science \cite{Restart in CS-1}
where the use of restart is now routine as it drastically improves
performance \cite{Restart in CS-1,Restart in CS-2,Restart in CS-3,Restart in CS-4,Restart in CS-5,Restart in CS-6}
of randomized algorithms \cite{Randomized algorithm-1,Randomized algorithm-2}.
The latter often display heavy-tailed run time distributions, and
diverging variances and even means \cite{Restart in CS-2,Restart in CS-3,Heavy-Tailed run times-1,Heavy-Tailed run times-2,Heavy-Tailed run times-3}.
Timely restart can then “censor” the tail of the run time distribution
and save the algorithm from getting “stuck” in sterile areas of the
search space where it is unlikely to find solutions. 

Restart is also relevant to many physical, chemical, and biological
processes as it is an integral part of the renowned Michaelis-Menten
Reaction Scheme (MMRS) \cite{MM}. In its original context the MMRS
depicts an enzyme which can reversibly bind a substrate to form a
complex. The substrate can then be converted by the enzyme to form
a product or, alternatively, unbind and restart the turnover cycle.
The MMRS has attracted interest for more than a century \cite{MMSI},
and today it is also used to describe heterogeneous catalysis \cite{Hetro1,Hetro2,Hetro3},
in vivo target search kinetics \cite{In Vivo Search1}, and other
processes. Two important predictions come from its classical analysis.
The rate of an enzymatic reaction should increase as the concentration
of the substrate increases and decrease as the unbinding rate increases
\cite{MM}. And yet, while the first prediction is well established
the second has never been tested experimentally.

Motivated by rapid advancements in single-molecule techniques \cite{Ever fluctuating,Fernandez1,Triggering enzymatic activity with force}
we scrutinized the role attributed to unbinding (restart) in Michaelis-Menten
reactions \cite{The Role}. We showed, via probabilistic---single-molecule
level---analysis, that unbinding of an enzyme from a substrate can
reduce the rate of product formation under some conditions, but that
it may also have an opposite effect. Indeed, as substrate concentrations
increase, a tipping point can be reached where an increase in the
unbinding rate results in an increase, rather than a decrease, of
the turnover rate. When this happens, a carefully chosen unbinding
rate can bring the enzymatic turnover rate to a maximum (mean FPT
to a minimum) by striking the right balance between the need to abort
prolonged reaction cycles and the need to avoid premature termination
of ongoing ones. Observations similar to ours were also made in the
context of search processes \cite{FD,Restart1,Restart2}, and a universal
condition for the existence of a non-vanishing optimal restart rate
was given in \cite{The Role}. It was clear, however, that the exact
identity of the latter may depend on fine details of the underlying
process (conversion of the substrate to a product in the case of enzymatic
reactions), and it thus seemed that little can be said in general
about optimal stochastic restart. 

In this letter, we address the question of universality in the fluctuations
of FPT processes subject to stochastic restart. As we have pointed
out in \cite{The Role}, any\emph{ }FPT process \cite{Redner}---be
it the time-to-target of a simple Brownian particle or that related
with a more sophisticated random searcher \cite{Search1,Search2,Search3}
or stochastic process \cite{Non-Brownian1,Non-Brownian2,FPT in complex invariant media,Geometry-controlled kinetics,Vibrational Shortcut,First Steps in Random Walks,ASIP1,ASIP2}---that
becomes subject to restart \cite{Restart1,Restart2,Restart3,Restart4,Restart5,Restart6,Restart7,Restart8,Restart9}
can naturally be described by the\textcolor{red}{{} }MMRS. This observation
has recently allowed us to give a unified treatment for the problem
of finding a restart rate which brings the mean completion time of
a generic process to a minimum (optimal restart) \cite{MMRS as a unified},
and it will be of value here again. We show that the relative fluctuation
in the FPT of an optimally restarted process is always unity. The
result is first illustrated, by means of example, on the now classical
problem of diffusion mediated search with stochastic restart \cite{Restart2}.
We then prove it in a more general setting and further demonstrate
its validity for a diverse set of examples. Before concluding, we
provide a probabilistic interpretation of our findings and generalize
the basic result to account for restart time overheads that inevitably
occur in many real life scenarios. It is only at this later stage
that we harness the full power of the MMRS. In what follows, we use\textbf{
}$\left\langle Z\right\rangle $,\textbf{ }$\sigma^{2}\left(Z\right)$
and $\tilde{Z}(s)\equiv\left\langle e^{-sZ}\right\rangle $ to denote,
respectively, the expectation, variance, and Laplace transform of
a real-valued random variable $Z$.

\begin{figure}[t]
\noindent \begin{centering}
\includegraphics[scale=0.45]{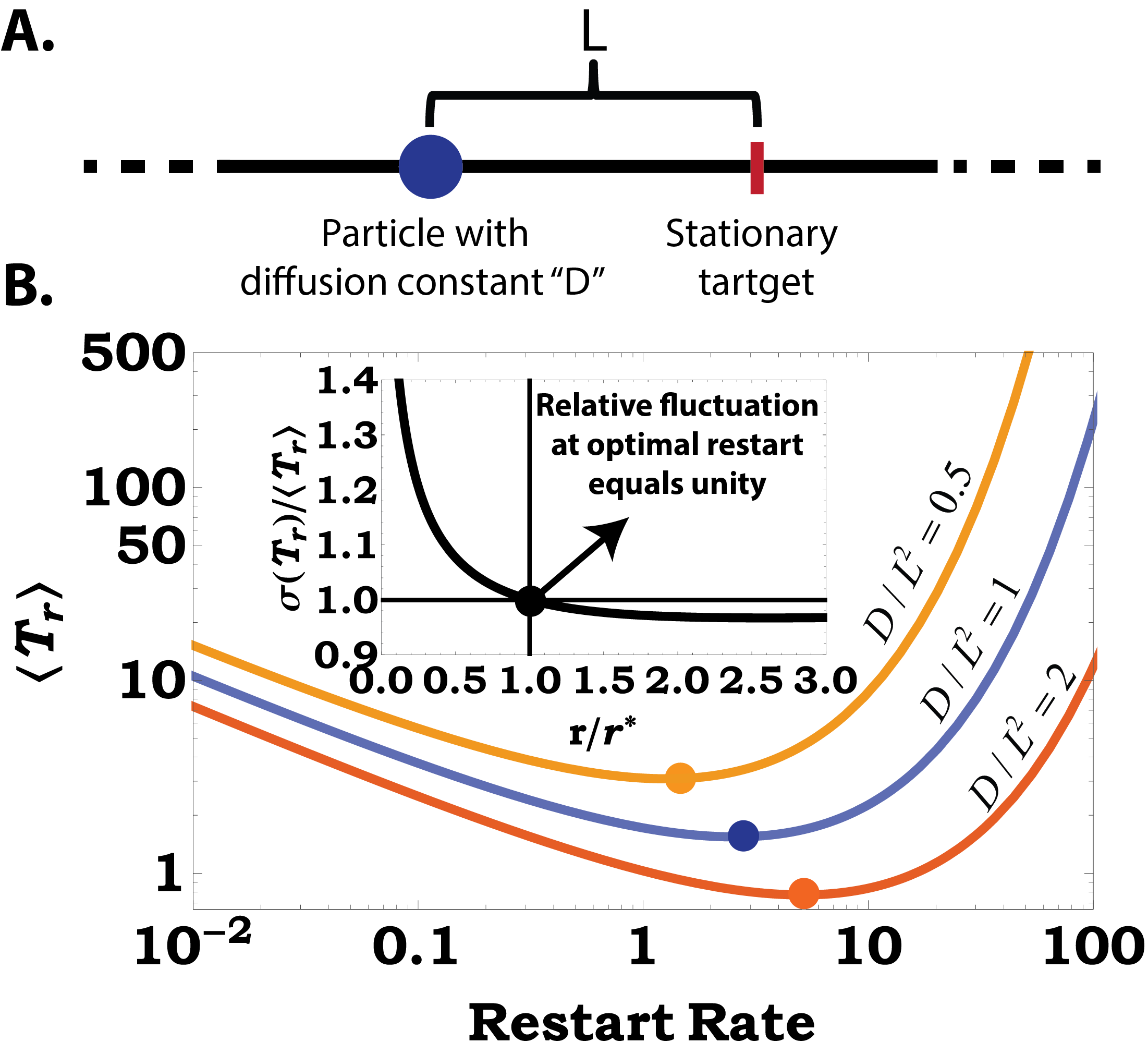}
\par\end{centering}

\protect\caption{\textbf{Color Online.} \textbf{A. }An illustration of diffusion mediated
search at time $t=0.$ \textbf{B. }Main. The mean FPT to target as
a function of the restart rate for different values of $D/L^{2}$.
The higher this ratio is, the higher the value of the optimal restart
rate $r^{*}$ which brings $\left\langle T_{r}\right\rangle $ to
a minimum (see positions marked with full circles). \textbf{Inset.
}The relative standard deviation in the FPT as a function of the restart
rate (normalized by the optimal restart rate). }
\end{figure}
\textbf{Diffusion with stochastic restart---a simple illustration
of a general principle. }Consider a particle ``searching'' for a
stationary target via one dimensional diffusion as is illustrated
in Fig. 1A. The particle starts at the origin, the initial distance
between the particle and the target is $L$, and the diffusion coefficient
of the particle is $D$. It has long been known that in this case
the mean FPT of the particle to the target diverges \cite{Redner,First Steps in Random Walks,elementary derivation}.
Consider now a scenario in which, on top of the above, the search
process is restarted, i.e., the particle is returned to its initial
position, at some given rate $r$. What is the mean FPT now? This
problem was studied in \cite{Restart2} by Evans \& Majumdar who found
that $T_{r}$, the FPT of the restarted process, is given by 

\begin{equation}
\left\langle T_{r}\right\rangle =\frac{e^{\sqrt{rL^{2}/D}}-1}{r}\,.\label{1}
\end{equation}
As anticipated, $\left\langle T_{r}\right\rangle $ depends on the
restart rate but it is interesting to note that it is finite for any
$r>0$. Moreover, an optimal restart rate which brings $\left\langle T_{r}\right\rangle $
to a minimum exists, as is illustrated in Fig. 1B, and one could readily
show that it is given by $r^{*}=(z^{*})^{2}D/L^{2}$, where $z^{*}\simeq1.59362...$
is the solution to $z/2=1-e^{-z}$. 

Evans \& Majumdar continued to compute the full distribution of $T_{r}$
and found that in Laplace space it is given by 
\begin{equation}
\tilde{T}_{r}(s)=\frac{s+r}{se^{\sqrt{(s+r)L^{2}/D}}+r}\,.\label{2}
\end{equation}
Moments could then be readily computed and we find that $\left\langle T_{r}^{2}\right\rangle =\frac{2-\sqrt{\frac{rL^{2}}{D}}e^{-\sqrt{rL^{2}/D}}-2e^{-\sqrt{rL^{2}/D}}}{r^{2}e^{-2\sqrt{rL^{2}/D}}}\,,$
from which it is easy to see that the relative standard deviation
in the completion time of the restarted processes is given by 
\begin{equation}
\frac{\sigma\left(T_{r}\right)}{\left\langle T_{r}\right\rangle }=\sqrt{\frac{e^{2\sqrt{rL^{2}/D}}-\sqrt{rL^{2}/D}e^{\sqrt{rL^{2}/D}}-1}{\left(e^{\sqrt{rL^{2}/D}}-1\right)^{2}}}\,.\begin{array}{l}
\text{ }\end{array}\label{3}
\end{equation}

The right hand side of Eq. (\ref{3}) has a form which suggests it
should be plotted as a function of $r/r^{*}$. We do so in the inset
of Fig. 1B only to find that when $r=r^{*}$ the relative fluctuation
in $T_{r}$ is exactly unity 
\begin{equation}
\frac{\sigma\left(T_{r^{*}}\right)}{\left\langle T_{r^{*}}\right\rangle }=1\,.\begin{array}{l}
\text{ }\end{array}\label{4}
\end{equation}
Quite strikingly, and as we will now show, the result in Eq. (\ref{4})
is not a peculiarity of diffusion but rather a universal property
common to all FPT processes subject to stochastic restart. 
\begin{figure}[t]
\noindent \centering{}\textbf{\includegraphics[scale=0.24]{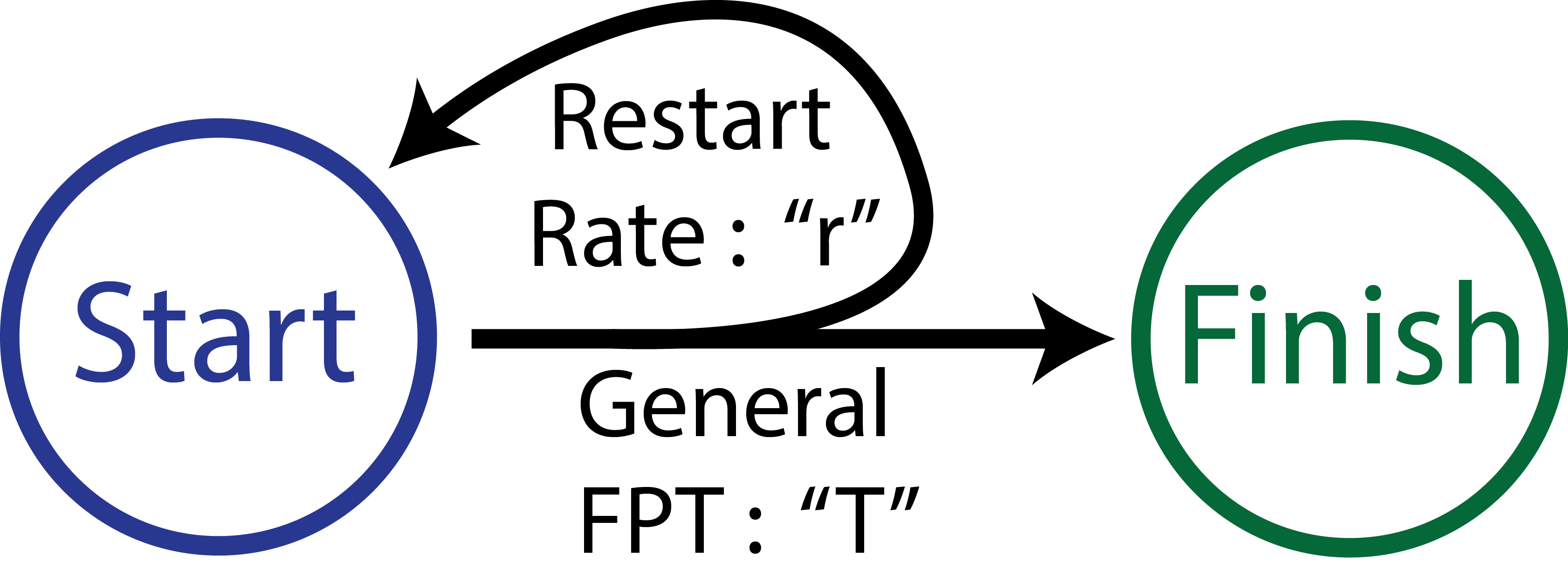}}\protect\caption{\textbf{Color Online.} An illustration of a generic process subject
to stochastic restart.}
\end{figure}

\textbf{Fluctuations in the first-passage-time of an optimally restarted
process are universal. }In deriving the main result of this paper
we consider the setting illustrated in Fig. 2. This setting captures
the model of diffusion with stochastic restart as a special case and
further allows us to generalize lessons learned from it. A generic
process starts at time zero and, if allowed to take place without
interruptions, will finish after a random time $T$. The process is,
however, restarted at some given rate $r$. Thus, if the process is
completed prior to restart the story there ends. Otherwise, the process
will start from scratch and begin completely anew. This procedure
repeats itself until the process reaches completion. 

Denoting the random completion time of the restarted process by $T_{r}$
it can then be seen that\textbf{ }
\begin{equation}
\begin{array}{l}
T_{r}=\left\{ \begin{array}{lll}
T &  & \text{if }T<R\text{ }\\
 & \text{ \ \ }\\
R+T_{r}^{\prime} &  & \text{if }R\leq T\text{ ,}
\end{array}\right.\\
\text{ }
\end{array}\label{5}
\end{equation}
where $R$ is an exponentially distributed random variable with rate
$r$ and $T_{r}^{\prime}$ an independent and identically distributed
copy of $T_{r}$. Taking the Laplace transform of $T_{r}$ we find
\begin{equation}
\tilde{T}_{r}(s)=\frac{\tilde{T}(s+r)}{\frac{s}{s+r}+\frac{r}{s+r}\tilde{T}(s+r)}\,,\label{6}
\end{equation}
and note that Eq. (\ref{6}) generalizes Eq. (\ref{2}). Indeed, in
the case of diffusion mediated search $\tilde{T}(s)=e^{-\sqrt{sL^{2}/D}}$
(Laplace space representation of the Lévy-Smirnov distribution \cite{First Steps in Random Walks})
and direct substitution of this expression into Eq. (\ref{6}) verifies
our claim. 
\begin{figure}
\noindent \begin{centering}
\includegraphics[scale=0.4]{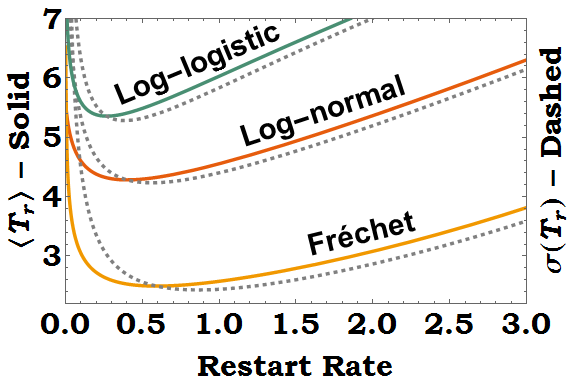}
\par\end{centering}

\protect\caption{\textbf{Color Online. }The mean (solid lines) and standard deviation
(dashed lines) in the FPT of a restarted process as a function of
the restart rate. Plots are made using Eq. (\ref{7}) for three different
time distributions of the underlying process subject to restart: (i)
Fréchet distribution $Pr(T\leq t)=e^{-t^{-\alpha}}\,\,(t>0)$, with
$\alpha=1$; (ii) Log-normal distribution $Pr(T\leq t)=\stackrel[0]{t}{\int}\left[x\sigma\sqrt{2\pi}\right]^{-1}\mathrm{exp}\left[-(\mathrm{ln}(x)-\mu)^{2}/2\sigma^{2}\right]dx\,\,(t>0)$,
with $\mu=1$ and $\sigma=1.2$; (iii) Log-logistic distribution $Pr(T\leq t)=\left[1+\left(t/\alpha\right)^{-\beta}\right]^{-1}\,\,(t>0)$,
with $\alpha=3.4$ and $\beta=1.45$. Equation (\ref{4}) asserts
that $\sigma\left(T_{r}\right)$ will cut $\left\langle T_{r}\right\rangle $
at the exact point at which the latter attains its minimum. }
\end{figure}

Using Eq. (\ref{6}) we compute the first two moments of $T_{r}$
\begin{equation}
\begin{array}{l}
\left\langle T_{r}\right\rangle =\frac{1-\tilde{T}(r)}{\tilde{T}(r)}\frac{1}{r}\,,\\
\\
\left\langle T_{r}^{2}\right\rangle =\frac{2\left(r\frac{d\tilde{T}(r)}{dr}-\tilde{T}(r)+1\right)}{r^{2}\tilde{T}(r)^{2}}\,.\\
\text{ }
\end{array}\label{7}
\end{equation}
Now, if the mean FPT of the restarted process attains a minimum (or
a maximum) at some\textcolor{red}{{} }$r^{*}>0$ we have $\frac{d\tilde{T}(r)}{dr}|_{r^{*}}=\frac{\tilde{T}(r^{*})(\tilde{T}(r^{*})-1)}{r^{*}}$
simply by taking the first derivative of $\left\langle T_{r}\right\rangle $
and equating it to zero at $r=r^{*}$. Substituting this result back
into Eq. (\ref{7}) gives $\left\langle T_{r^{*}}^{2}\right\rangle =2\left(\tilde{T}^{2}(r^{*})-2\tilde{T}(r^{*})+1\right)/\left(\left(r^{*}\right)^{2}\tilde{T}(r^{*})^{2}\right),$
from which it is easy to see that $\sigma^{2}\left(T_{r^{*}}\right)=\left(\tilde{T}(r^{*})-1\right)^{2}/\left(\left(r^{*}\right)^{2}\tilde{T}(r^{*})^{2}\right).$
Comparing this result with Eq. (\ref{7}) we conclude that Eq. (\ref{4})
holds for an arbitrary FPT process (also see illustration in Fig.
3). 

Equation (\ref{4}) has an interesting probabilistic interpretation.
Examining Fig. 2, one could ask what determines whether $\left\langle T\right\rangle $
is larger or smaller than $\left\langle T_{\delta r}\right\rangle $
for an infinitesimal $\delta r$? This question can be answered by
letting the original process repeat itself over and over again without
restarting it. Visiting this process at a random point in time one
could ask what is the mean time left till the next completion event
occurs? This time is known as the mean residual life time of the process
and is given by $\left\langle T_{res}\right\rangle =\frac{1}{2}\left\langle T^{2}\right\rangle /\left\langle T\right\rangle $
\cite{Gallager}. It is intuitively clear that $\frac{d\left\langle T_{r}\right\rangle }{dr}|_{r=0}$
will be negative (positive) whenever $\left\langle T_{res}\right\rangle $
is larger (smaller) than $\left\langle T\right\rangle $. Applying
the same logic to a process which is already subject to restart (simply
by seeing it as an original process of itself), we conclude that when
a process is restarted at an optimal rate\textcolor{red}{{} }its mean
and mean residual life time must be equal. Indeed, any deviation from
equality is in contraindication to optimality as it implies $\frac{d\left\langle T_{r}\right\rangle }{dr}|_{r^{*}}\neq0$.
We thus have $\frac{1}{2}\left\langle T_{r^{*}}^{2}\right\rangle /\left\langle T_{r^{*}}\right\rangle =\left\langle T_{r^{*}}\right\rangle $
from which Eq. (\ref{4}) follows immediately. 

Recapitulating this section we note that the normalized completion
time of an optimally restarted process $T_{r^{*}}/\left\langle T_{r^{*}}\right\rangle $
has, by definition, mean one and, as we have just shown, a standard
deviation which also equals unity. It thus follows that $\left\langle e^{-sT_{r^{*}}/\left\langle T_{r^{*}}\right\rangle }\right\rangle \simeq1-s+s^{2}+o(s^{2})$---a
form which to second order coincides with the Laplace transform of
the exponential distribution $\left(1+s\right)^{-1}$. However, and
as we illustrate in Fig. 4, the distribution of $T_{r^{*}}/\left\langle T_{r^{*}}\right\rangle $
is not universal and deviations from exponentiality may arise for
$s\gg1$.

\begin{figure}
\noindent \begin{centering}
\includegraphics[scale=0.17]{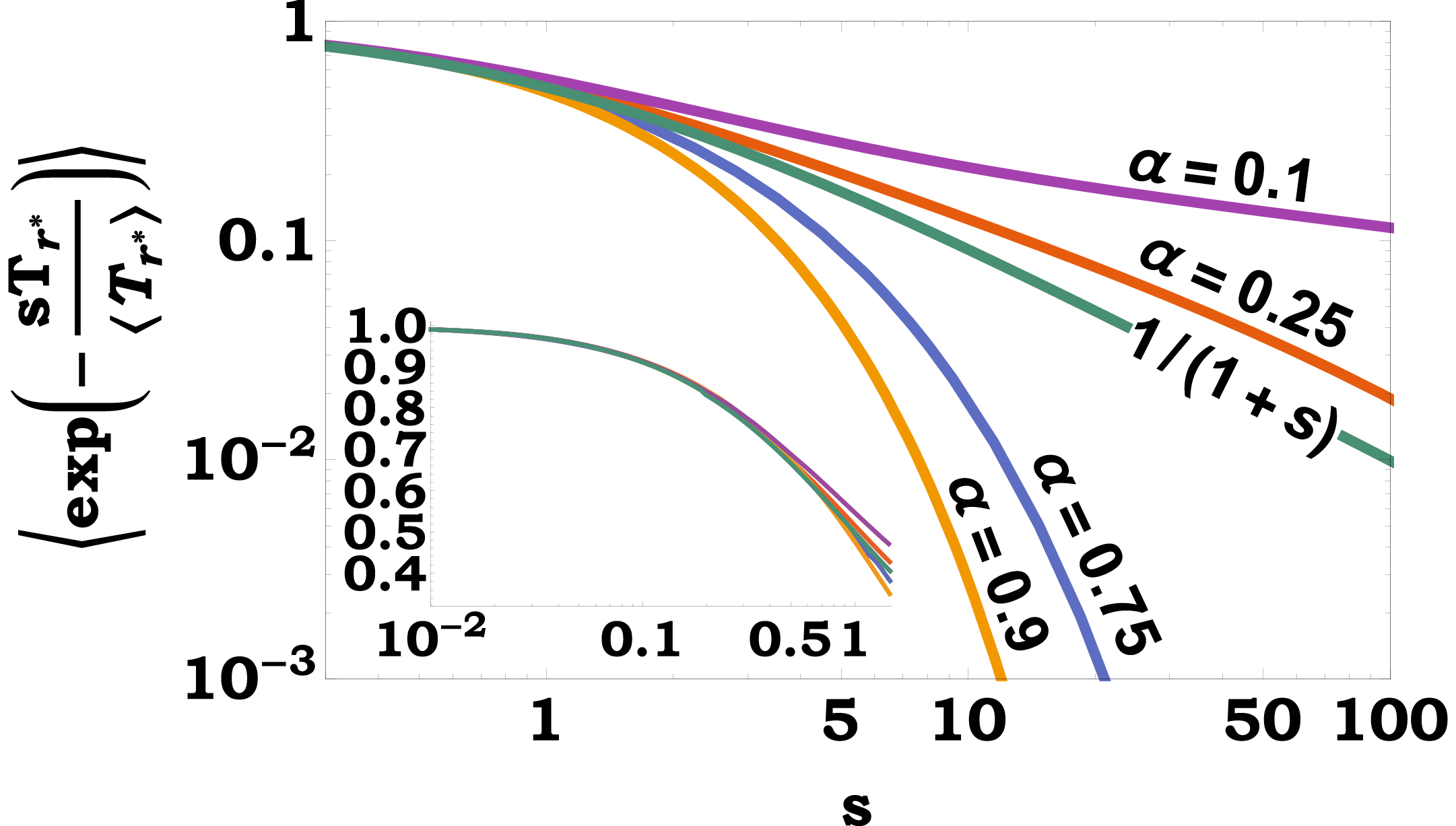}
\par\end{centering}

\protect\caption{\textbf{Color Online. }The distribution of $T_{r^{*}}/\left\langle T_{r^{*}}\right\rangle $
is non-universal. Consider a generalization of diffusion with stochastic
restart. Letting $\tilde{T}(s)=e^{-\left(\tau s\right){}^{\alpha}}$
we note that for $\tau=L^{2}/D$ and $\alpha=1/2$ we have the Lévy-Smirnov
distribution discussed above, and in general for $0<\alpha<1$ the
Laplace transform of the one sided Lévy distribution. In \cite{MMRS as a unified}
we showed that the optimal restart rate for this distribution is given
by $r^{*}=(z^{*})^{1/\alpha}/\tau$, where $z^{*}$ is the solution
to $\alpha z=1-e^{-z}$. This now allows us to compute $\left\langle T_{r^{*}}\right\rangle $
and plot $\left\langle e^{-sT_{r^{*}}/\left\langle T_{r^{*}}\right\rangle }\right\rangle $
for different values of $\alpha$ while fixing $\tau=1$ for convenience.
It is clearly visible that while all curves fall on top of each other,
and on top of $\left(1+s\right)^{-1}$, for $s\ll1$ their behavior
for $s\gg1$ depends on the value of $\alpha$. }
\end{figure}
\textbf{Stochastic restart with time overheads. }Deriving Eq. (\ref{4})
we have implicitly assumed that restart does not bear with it any
time penalty. And yet, when a computer program is stopped restarting
it may involve a time overhead. Similarly, when an enzyme unbinds
from its substrate time will pass before it binds a new one. This
type of complication can be addressed by generalizing the stochastic
renewal law in Eq. (\ref{5}) to read 

\begin{equation}
T_{r}=T_{on}+\left\{ \begin{array}{lll}
T &  & \text{if }T<R\text{ }\\
 & \text{ \ \ }\\
R+T_{r}^{\prime} &  & \text{if }R\leq T\text{ ,}
\end{array}\right.\label{8}
\end{equation}
where $T_{on}$ is a generally distributed random time which collectively
accounts for ``delays'' that may arise prior to any completion attempt.
Equation (\ref{8}) furnishes a mathematical description of the MMRS
(see introduction to this letter) and the effect restart has on $\left\langle T_{r}\right\rangle $
in this case was extensively explored in \cite{The Role,MMRS as a unified}.
Here, we will be interested in the effect it has on fluctuations.

Utilizing Eq. (\ref{8}) one could show that 
\begin{equation}
\tilde{T}_{r}(s)=\frac{\tilde{T}(s+r)\tilde{T}_{on}(s)}{1+\frac{r}{s+r}\tilde{T}_{on}(s)\left(\tilde{T}(s+r)-1\right)}\,,\label{9}
\end{equation}
and we find that $\left\langle T_{r}\right\rangle =\frac{r\left\langle T_{on}\right\rangle +1-\tilde{T}(r)}{r\tilde{T}(r)}$
and $\left\langle T_{r}^{2}\right\rangle =\frac{2\left(1-\tilde{T}(r)\right)\left(1+r\left\langle T_{on}\right\rangle \right)^{2}+2r\left(1+r\left\langle T_{on}\right\rangle \right)\frac{d\tilde{T}(r)}{dr}+r^{2}\tilde{T}(r)\left\langle T_{on}^{2}\right\rangle }{r^{2}\tilde{T}(r)^{2}}$.
Moreover, in \cite{MMRS as a unified} we showed that if $\left\langle T_{r}\right\rangle $
receives a minimum (or a maximum) at some $r^{*}>0$ the following
equation must hold $\frac{\tilde{T}(r^{*})(\tilde{T}(r^{*})-1)}{\left(r^{*}\right)^{2}\frac{d\tilde{T}(r)}{dr}|_{r^{*}}}-\frac{1}{_{r*}}=\left\langle T_{on}\right\rangle $.
Solving for $\frac{d\tilde{T}(r)}{dr}|_{r^{*}}$ and substituting
the result in the expression for $\left\langle T_{r^{*}}^{2}\right\rangle $
we obtain (see Fig. 5 for illustration) 
\begin{equation}
\frac{\sigma\left(T_{r^{*}}\right)}{\left\langle T_{r^{*}}\right\rangle }=\sqrt{1+\frac{\sigma^{2}\left(T_{on}\right)-\left\langle T_{on}\right\rangle ^{2}}{\tilde{T}(r^{*})\left\langle T_{r^{*}}\right\rangle ^{2}}}\,.\begin{array}{l}
\text{ }\end{array}\label{10}
\end{equation}
Equation (\ref{10}) generalizes Eq. (\ref{4}) and while the result
is no longer universal \cite{Not universal} it is surprisingly elegant
. 

\begin{figure}
\noindent \begin{centering}
\includegraphics[scale=0.17]{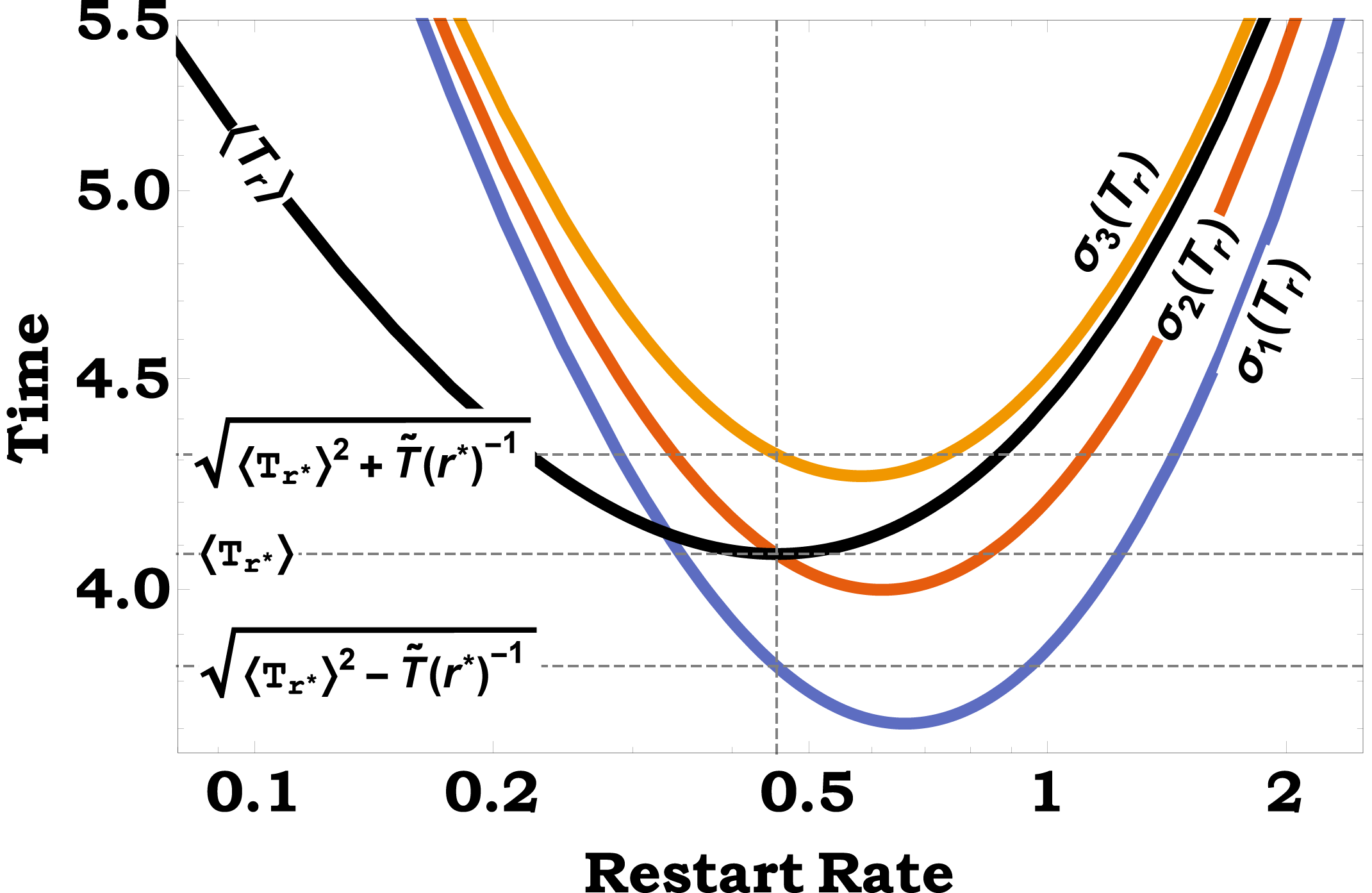}
\par\end{centering}

\protect\caption{\textbf{Color Online. }The mean and standard deviation of $T_{r}$
for three different cases of diffusion mediated search with stochastic
restart and time overheads. In all three cases $\tilde{T}(s)=e^{-\sqrt{s}}$
($L^{2}/D=1$), $\left\langle T_{on}\right\rangle =1$, and $\left\langle T_{r}\right\rangle $
is hence one and the same. The standard deviations $\sigma_{i}(T_{r})$
are, however, quite distinct since here we chose $\sigma_{i}(T_{on})=i-1$
~for~ $i=1,\,2,\,3$. These values could, for example, be associated
with sharp (deterministic), exponential, and Gamma distributed delays
(in this order). Note that in all three cases $\sigma_{i}\left(T_{r^{*}}\right)=\sqrt{\left\langle T_{r^{*}}\right\rangle ^{2}+\frac{\sigma_{i}^{2}\left(T_{on}\right)-\left\langle T_{on}\right\rangle ^{2}}{\tilde{T}(r^{*})}}$
in accord with Eq. (\ref{10}).}
\end{figure}
\textbf{Conclusions.}\textcolor{red}{{} }The advent of novel single
molecule and cell techniques has truly invigorated the experimental
\cite{FluctuationsExp1,FluctuationsExp2,FluctuationsExp3,FluctuationsExp4,FluctuationsExp5,FluctuationsExp6,FluctuationsExp7}
and theoretical \cite{FluctuationsT0,FluctuationsT1,FluctuationsT2,FluctuationsT3,FluctuationsT4,FluctuationsT5,FluctuationsT6,FluctuationsT7}
study of fluctuation phenomena. Notable in that regard are studies
directed towards questions of universality as they allow us to draw
broad, model independent, conclusions which are in turn widely applicable
\cite{Universality1,Universality2,Universality3,FPT in complex invariant media,Geometry-controlled kinetics,Universality4}.
In this letter we studied the effect of stochastic restart on the
fluctuations of a generic process and showed that when the restart
rate is optimal, in the sense that it minimizes (or maximizes) the
mean FPT of the process, fluctuations are universal. The prevalence
of FPT processes in the sciences and the multitude of perspectives
which bring one to consider (optimal) restart encourage us to think
that the results we have obtained will be of general use. Applications
to the field of single molecule enzymology are particularly interesting
since enzymes are subject to selective pressure which may have dialed
unbinding (restart) rates optimal. Moreover, note that whenever $T_{on}$
arises from the amalgamation of many, independent and low intensity,
events it will admit Poisson statistics for which $\sigma\left(T_{on}\right)=\left\langle T_{on}\right\rangle $.
In this case Eq. (\ref{10}) reduces to Eq. (\ref{4}) and the common
scenario in which numerous substrate molecules try to bind an enzyme
is a good example for a situation of that sort. Will it be found that
selective pressure towards optimal restart rendered fluctuations in
enzymatic turnover rates universal, it would not be the first time
that proteins are found occupying a very special niche within a vastly
accessible phase space \cite{Proteins1,Proteins2}, and yet another
example for the importance of optimality and extremality ideas in
Biological Physics \cite{Optimality}.

\textbf{Acknowledgments.} Shlomi Reuveni gratefully acknowledges support
from the James S. McDonnell Foundation via its postdoctoral fellowship
in studying complex systems. We thank Ori Hirschberg and Ariel Amir
for reading the paper and for providing comments. We thank Cristopher
Moore for turning our attention to applications in computer science
and for referring us to relevant literature.


\begin{thebibliography}{10}
{\small{}\bibitem{Restart in CS-1}M. Luby, A. Sinclair and D. Zuckerman,
}\emph{\small{}Inform. Process. Lett.}{\small{} }\textbf{\small{}47}{\small{}
(4), 173-180, (1993). }{\small \par}

{\small{}\bibitem{Restart in CS-2}C.P. Gomes, B. Selman and H. Kautz,
}\emph{\small{}AAAI/IAAI}{\small{} }\textbf{\small{}98}{\small{},
431, (1998).}{\small \par}

{\small{}\bibitem{Restart in CS-3}T. Walsh, }\emph{\small{}IJCAI}{\small{},
}\textbf{\small{}99}{\small{}, (1999).}{\small \par}

{\small{}\bibitem{Restart in CS-4}M. W. Moskewicz et.al., In Proceedings
of the 38th annual Design Automation Conference, 530-535, ACM, (2001).}{\small \par}

{\small{}\bibitem{Restart in CS-5}A. Montanari and R. Zecchina, }\emph{\small{}Phys.
Rev. Lett.,}{\small{} }\textbf{\small{}88}{\small{}, 178701, (2002). }{\small \par}

{\small{}\bibitem{Restart in CS-6}J. Huang, }\emph{\small{}In IJCAI}{\small{},
}\textbf{\small{}7}{\small{}, 2318-2323, (2007).}{\small \par}

{\small{}\bibitem{Randomized algorithm-1}L. Lovasz, Random walks
on graphs: A survey in Combinatronics (Bolyai Society for Mathematical
Studies, 1996), Vol. 2, p. 1.}{\small \par}

{\small{}\bibitem{Randomized algorithm-2}C. Moore \& S. Mertens,
The Nature of Computation, Oxford University Press, (2011).}{\small \par}

{\small{}\bibitem{Heavy-Tailed run times-1}C. P. Gomes, B. Selman
and N. Crato. In Principles and Practice of Constraint Programming-CP
1997 (pp. 121-135). Springer Berlin Heidelberg.}{\small \par}

{\small{}\bibitem{Heavy-Tailed run times-2}H. Chen, C. P. Gomes and
B. Selman, In Principles and Practice of Constraint Programming—CP
2001 (pp. 408-421). Springer Berlin Heidelberg.}{\small \par}

{\small{}\bibitem{Heavy-Tailed run times-3}H. Jia \& C. Moore, In
Principles and Practice of Constraint Programming–CP 2004 (pp. 742-746).
Springer Berlin Heidelberg.}{\small \par}

{\small{}\bibitem{MM}L. Menten \& M.I. Michaelis, }\textit{\small{}Biochem
Z}{\small{}, }\textbf{\small{}49}{\small{}, 333, (1913).}{\small \par}

{\small{}\bibitem{MMSI}A. Cornish-Bowden \& C. P. Whitman (Editors),
A century of Michaelis - Menten kinetics, }\emph{\small{}FEBS Letters}{\small{},
}\textbf{\small{}587}{\small{}, 17, 2711, (2013). }{\small \par}

{\small{}\bibitem{Hetro1}M. B. J. Roeffaers }\emph{\small{}et.al.}{\small{},
}\emph{\small{}Nature}{\small{}, }\textbf{\small{}439}{\small{}, 572,
(2006).}{\small \par}

{\small{}\bibitem{Hetro2}W. Xu }\emph{\small{}et.al.}{\small{}, }\emph{\small{}Nat.
Mater.}{\small{},}\textbf{\small{} 7}{\small{}, 992, (2008).}{\small \par}

{\small{}\bibitem{Hetro3}K. P. F. Janssen }\emph{\small{}et.al.}{\small{},
}\emph{\small{}Chem. Soc. Rev.}{\small{}, }\textbf{\small{}43}{\small{},
990, (2014).}{\small \par}

{\small{}\bibitem{In Vivo Search1}J. Fei }\emph{\small{}et. al.}{\small{},
Science, }\textbf{\small{}347}{\small{}, 6228, (2015).}{\small \par}

{\small{}\bibitem{Ever fluctuating}B.P. English }\emph{\small{}et.
al.}{\small{}, }\emph{\small{}Nature Chem. Biol}{\small{}., }\textbf{\small{}2}{\small{},
87-94, (2006). }{\small \par}

{\small{}\bibitem{Fernandez1}A.P. Wiita }\emph{\small{}et. al.}{\small{},
}\emph{\small{}Nature}{\small{}, }\textbf{\small{}450}{\small{}, 7166,
124, (2007). }{\small \par}

{\small{}\bibitem{Triggering enzymatic activity with force}H. Gumpp
}\emph{\small{}et. al.}{\small{}, }\emph{\small{}Nano Lett.,}{\small{}
}\textbf{\small{}9}{\small{}, 3290–3295, (2009). }{\small \par}

{\small{}\bibitem{The Role}S. Reuveni }\emph{\small{}et. al.}{\small{},
}\emph{\small{}Proc. Natl. Acad. Sci. U.S.A.,}{\small{} }\textbf{\small{}111}{\small{},
(12), 4391, (2014).}{\small \par}

{\small{}\bibitem{FD}P. H. von Hippel \& O. G. Berg., }\textit{\small{}J.
Biol. Chem.}{\small{}, }\textbf{\small{}264}{\small{}, 675, (1989).}{\small \par}

\bibitem{Restart1}{\small{}I. Eliazar }\emph{\small{}et. al.}{\small{},
}\emph{\small{}J. Phys}{\small{}, }\emph{\small{}Condens. Matter.}{\small{}
}\textbf{\small{}19}{\small{}, 065140, (2007).}{\small \par}

\bibitem{Restart2}{\small{}M.R. Evans \& S.N. Majumdar, }\textit{\small{}Phys.
Rev. Lett.}{\small{},}\textbf{\small{} 106}{\small{}, 160601, (2011).}{\small \par}

{\small{}\bibitem{Redner}S. Redner, }\emph{\small{}A Guide to First-Passage
Processes,}{\small{} Cambridge University Press, (2001).}{\small \par}

{\small{}\bibitem{Search1}M. A. Lomholt et.al., }\emph{\small{}Proc.
Natl. Acad. Sci. U.S.A.}{\small{}, }\textbf{\small{}105}{\small{}
(32) 11055, (2008).}{\small \par}

{\small{}\bibitem{Search2}O. Bénichou et.al., }\emph{\small{}Rev.
Mod. Phys.,}{\small{} }\textbf{\small{}83}{\small{}, 81, (2011).}{\small \par}

{\small{}\bibitem{Search3}V. V. Palyulin et.al., }\emph{\small{}Proc.
Natl. Acad. Sci. U.S.A.,}{\small{} }\textbf{\small{}111}{\small{}
(8) 2931, (2014).}{\small \par}

{\small{}\bibitem{Non-Brownian1}R. Metzler \& J. Klafter, }\emph{\small{}Physics
Reports,}{\small{} }\textbf{\small{}339}{\small{}, 1, (2000).}{\small \par}

{\small{}\bibitem{Non-Brownian2}I.M. Sokolov et.al., }\emph{\small{}Physics
Today,}{\small{} }\textbf{\small{}55}{\small{}, 11, 48, (2002). }{\small \par}

{\small{}\bibitem{FPT in complex invariant media}S. Condamin, et.al.,
}\emph{\small{}Nature,}{\small{} }\textbf{\small{}450}{\small{}, 77-
80, (2007).}{\small \par}

{\small{}\bibitem{Geometry-controlled kinetics}O. Bénichou, et.al.,
}\emph{\small{}Nature Chemistry,}{\small{} }\textbf{\small{}2}{\small{},
472 (2010).}{\small \par}

{\small{}\bibitem{Vibrational Shortcut}S. Reuveni, R. Granek and
J. Klafter, }\emph{\small{}Phys. Rev. E,}{\small{} }\textbf{\small{}81}{\small{},
040103(R), (2010).}{\small \par}

{\small{}\bibitem{First Steps in Random Walks}J. Klafter \& I.M.
Sokolov, First Steps in Random Walks, Oxford University Press, (2011).}{\small \par}

{\small{}\bibitem{ASIP1}S. Reuveni, I. Eliazar and U. Yechiali, }\emph{\small{}Phys.
Rev. E,}{\small{} }\textbf{\small{}84}{\small{}, 041101, (2011). }{\small \par}

{\small{}\bibitem{ASIP2}S. Reuveni, I. Eliazar and U. Yechiali, }\emph{\small{}Phys.
Rev. Lett.,}{\small{} }\textbf{\small{}109}{\small{}, 020603, (2012).}{\small \par}

{\small{}\bibitem{Restart3}M.R. Evans \& S.N. Majumdar, }\emph{\small{}J.
Phys. A: Math. Theor.,}{\small{} }\textbf{\small{}44}{\small{}, 435001,
(2011).}{\small \par}

{\small{}\bibitem{Restart4}L. Kusmierz et. al., }\emph{\small{}Phys.
Rev. Lett.,}{\small{} }\textbf{\small{}113}{\small{}, 220602, (2014).}{\small \par}

{\small{}\bibitem{Restart5}S. C. Manrubia \& D. H. Zanette}\textit{\small{},
Phys. Rev. E, }\textbf{\textit{\small{}59}}\textit{\small{}, 4945,
(1999).}{\small \par}

{\small{}\bibitem{Restart6}M. Montero \& J. Villarroel, }\textit{\small{}Phys.
Rev. E,}{\small{} }\textbf{\small{}87}{\small{}, 012116, (2013).}{\small \par}

\bibitem{Restart7}{\small{}S. N. Majumdar, S. Sabhapandit and G.
Schehr. }\emph{\small{}Phys. Rev. E,}{\small{} }\textbf{\small{}92}{\small{},
052126, (2015).}{\small \par}

{\small{}\bibitem{Restart8}Ł. Kuśmierz and E. Gudowska-Nowak, }\emph{\small{}Phys.
Rev. E,}{\small{} }\textbf{\small{}92}{\small{}, 052127, (2015).}{\small \par}

{\small{}\bibitem{Restart9}S. Eule \& J. Metzger, arXiv:1510.07876.}{\small \par}

{\small{}\bibitem{MMRS as a unified}T. Rotbart, S. Reuveni and M.
Urbakh, arXiv:1509.05071, to be published in Phys. Rev. E. }{\small \par}

{\small{}\bibitem{elementary derivation}Also see S. Kostinski \&
A. Amir, arXiv:1509.04800, for a pedagogical derivation of a related
result. }{\small \par}

{\small{}\bibitem{Gallager}See definition and formula of the time-averaged
residual life time of a renewal process in: R. G. Gallager, Stochastic
Processes: Theory for Applications, Cambridge University Press, (2013). }{\small \par}

{\small{}\bibitem{Not universal}Note dependence on $\tilde{T}(r^{*})$
and hence on the details of the restarted process.}{\small \par}

{\small{}\bibitem{FluctuationsExp1}H.P. Lu et.al., }\emph{\small{}Science}{\small{},
}\textbf{\small{}282}{\small{}, 1877, (1998).}{\small \par}

{\small{}\bibitem{FluctuationsExp2}L. Edman }\emph{\small{}et. al.}{\small{},
}\emph{\small{}Chem. Phys.}{\small{} }\textbf{\small{}247}{\small{},
11, (1999). }{\small \par}

{\small{}\bibitem{FluctuationsExp3}M. B. Elowitz, et al. }\emph{\small{}Science,}{\small{}
}\textbf{\small{}297}{\small{}, 5584, 1183-1186, (2002).}{\small \par}

{\small{}\bibitem{FluctuationsExp4}O. Flomenbom }\emph{\small{}et.
al.}{\small{}, }\emph{\small{}Proc. Natl. Acad. Sci. U.S.A.}{\small{},
}\textbf{\small{}102}{\small{}, 2368, (2005).}{\small \par}

{\small{}\bibitem{FluctuationsExp5}Y. Taniguchi et. al. }\emph{\small{}Science,}{\small{}
}\textbf{\small{}329}{\small{}, 5991, 533-538, (2010).}{\small \par}

{\small{}\bibitem{FluctuationsExp6}T.M. Norman, N.D. Lord, J. Paulsson,
R. Losick, }\emph{\small{}Nature,}\textbf{\small{} 503}{\small{},
(7477), 481, (2013).}{\small \par}

{\small{}\bibitem{FluctuationsExp7}O. Sandler et. al., }\emph{\small{}Nature,}{\small{}
}\textbf{\small{}519}{\small{}, 468–471, (2015).}{\small \par}

{\small{}\bibitem{FluctuationsT0}J. Paulsson. }\emph{\small{}Nature,}{\small{}
}\textbf{\small{}427}{\small{}, 415-418, (2004).}{\small \par}

{\small{}\bibitem{FluctuationsT1}S.C. Kou }\emph{\small{}et. al.}{\small{},
}\emph{\small{}J. Phys. Chem. B,}{\small{} }\textbf{\small{}109}{\small{},
19068, (2005).}{\small \par}

{\small{}\bibitem{FluctuationsT2}S. Reuveni, R. Granek and J. Klafter,
}\emph{\small{}PNAS,}{\small{} }\textbf{\small{}107}{\small{} (31),
13696, (2010).}{\small \par}

{\small{}\bibitem{FluctuationsT3}S. Yang }\emph{\small{}et. al.}{\small{},
}\emph{\small{}Biophysical Journal,}{\small{} }\textbf{\small{}101}{\small{},
519, (2011). }{\small \par}

{\small{}\bibitem{FluctuationsT4}A. Hilfinger \& J. Paulsson, }\emph{\small{}Proc.
Natl. Acad. Sci. U.S.A.,}{\small{} }\textbf{\small{}109}{\small{}(29),
12167-72 (2011).}{\small \par}

{\small{}\bibitem{FluctuationsT5}S. Reuveni, J. Klafter and R. Granek,
}\emph{\small{}Phys. Rev. Lett.,}{\small{} }\textbf{\small{}108}{\small{},
068101, (2012).}{\small \par}

{\small{}\bibitem{FluctuationsT6}S. Reuveni, J. Klafter and R. Granek,
}\emph{\small{}Phys. Rev. E,}{\small{} }\textbf{\small{}85}{\small{},
011906, (2012).}{\small \par}

{\small{}\bibitem{FluctuationsT7}A. Amir, }\emph{\small{}Phys. Rev.
Lett., }\textbf{\small{}112}{\small{}, 20, 208102, (2014).}{\small \par}

{\small{}\bibitem{Universality0}I. Lestas, G. Vinnicombe and J. Paulsson,
}\emph{\small{}Nature,}{\small{} }\textbf{\small{}467}{\small{}, (7312),
163-4 (2010).}{\small \par}

{\small{}\bibitem{Universality1}S. Chaudhury, J. Cao, and N. A. Sinitsyn,
}\emph{\small{}J. Phys. Chem. B,}{\small{} }\textbf{\small{}117}{\small{},
503 (2013).}{\small \par}

{\small{}\bibitem{Universality2}J.R. Moffitt \& C. Bustamante, }\emph{\small{}FEBS
J.,}{\small{} }\textbf{\small{}281 }{\small{}(2), 498, (2014).}{\small \par}

{\small{}\bibitem{Universality3}A. C. Barato \& Udo Seifert, }\emph{\small{}Phys.
Rev. Lett.,}{\small{} }\textbf{\small{}115}{\small{}, 18, 188103,
(2015).}{\small \par}

{\small{}\bibitem{Universality4}M. Chupeau, O. Bénichou and R. Voituriez,
}\emph{\small{}Nature Physics,}{\small{} }\textbf{\small{}11}{\small{},
844–847, (2015).}{\small \par}

{\small{}\bibitem{Proteins1}S. Reuveni, R. Granek and J. Klafter,
}\emph{\small{}Phys. Rev. Lett.,}{\small{} }\textbf{\small{}100}{\small{},
208101, (2008).}{\small \par}

{\small{}\bibitem{Proteins2}M. de Leeuw, S. Reuveni, J. Klafter and
R. Granek, }\emph{\small{}PLoS ONE,}{\small{} }\textbf{\small{}4}{\small{}(10),
(2009).}{\small \par}

{\small{}\bibitem{Optimality}Bialek, William. Biophysics: searching
for principles. Princeton University Press, (2012).}\end{thebibliography}
\end{document}